\documentclass[fleqn,usenatbib]{mnras}

\usepackage[T1]{fontenc}
\usepackage{pdflscape}

\DeclareRobustCommand{\VAN}[3]{#2}
\let\VANthebibliography\thebibliography
\def\thebibliography{\DeclareRobustCommand{\VAN}[3]{##3}\VANthebibliography}

\usepackage{graphicx}	
\usepackage{amsmath}	
\usepackage{amssymb}	
\usepackage{arydshln}
\usepackage{multirow}

\usepackage{gensymb}

\usepackage{newtxtext,newtxmath}

\title[How to mis-align your bow shock]{Why do massive stars form bow shocks? \\ \textit{Bulk ISM motion as the main driver of bow shock formation and geometry}}

\author[Van den Eijnden \& Kaper]{Jakob van den Eijnden$^{1}$\thanks{a.j.vandeneijnden@uva.nl} and Lex Kaper$^{1}$
\\
$^{1}$Anton Pannekoek Institute for Astronomy, Universiteit van Amsterdam, Science Park 904, 1098, XH, Amsterdam, The Netherlands\\
}

\date{Accepted XXX. Received YYY; in original form ZZZ}

\pubyear{2026}

\begin{document}
\label{firstpage}
\pagerange{\pageref{firstpage}--\pageref{lastpage}}
\maketitle

\begin{abstract}
Bow shocks are one the most commonly observed impact sites of massive star feedback and are typically interpreted as evidence for supersonic stellar motion. However, some bow shocks have been reported around sub-sonic massive stars or orientated differently than the stellar direction of movement, suggesting a substantial influence of the interstellar medium (ISM). Here, we quantitatively investigate the dominant cause of bow shock formation around massive stars, combining infrared bow shock samples with Gaia DR3. We find that only half ($52\pm2\%$) of the bow shocks is driven by a supersonic star and that $70\pm2$\% is substantially misaligned ($>30\degree$). Only $21\pm2\%$ of the studies systems can be regarded as \textit{classical} bow shocks, driven by a supersonic star in the direction of stellar movement. For the remaining $79\pm2\%$ of systems, the role of bulk ISM movement in creating the shock is equal to or dominant over the stellar movement. We also introduce a statistical method to quantify the ISM motions required to play this dominant role. We find that random ISM motions at $10-15$ km s$^{-1}$ suffice to drive the existence and orientations of massive star bow shocks. For systems facing HII regions, we find evidence that outflows at $>25$ km s$^{-1}$ drive the nearby shocks. We conclude that ISM motion is the dominant factor in creating massive star bow shocks and discuss how ignoring these motions can lead to underestimates of the stellar wind's mass loss rate by an order of magnitude. 
\end{abstract}

\begin{keywords}
stars: massive -- stars: mass-loss -- stars: winds, outflows -- stars: kinematics and dynamics -- ISM: kinematics and dynamics -- shock waves
\end{keywords}

\section{Introduction}

Despite being short-lived and forming relatively rarely, massive stars (e.g., early-type stars, mass $>8M _\odot$) are key drivers of feedback processes on scales from their local environments to their host galaxies. These massive stars, individually, shape their direct surroundings through a combination of their bright radiation and stellar winds \citep{2014ARA&A..52..487S,vink2022}. Collectively, in young stellar clusters, feedback of massive stars can drive large scale outflows that regulate the star-formation rate and thereby drive the evolution of the host galaxy \citep{2012MNRAS.421.3522H,2022NatAs...6..647C}. Through wide-ranging avenues -- radiative and mechanical feedback, supernova explosions, the formation of (accreting) compact objects -- massive stars play a key role in the origin and acceleration of Galactic cosmic rays \citep[e.g.,][]{2013MNRAS.431..415B,Aharonian2019,2024Natur.634..557A}.

Feedback sites around individual massive stars can be challenging to observe directly. Supernova remnants are an obvious and common example, detected in large numbers up to extragalactic environments. However, during earlier stages of massive star evolution, examples are more scarce. In the Milky Way, feedback bubbles have been detected around some Wolf-Rayet stars \citep{Prajapati2019}. Ejecta and shells have also been observed around Luminous Blue Variables (famously visible in, e.g., the ISM structure around $\eta$ Car; \citealt{2001MNRAS.327...46B}, \citealt{2002MNRAS.337.1252S}; see also \citealt{gvaramadze2010}) and around sites of binary stellar mergers \citep{mahy2017,frost2024}. Whether these sites are detected, depends on both the massive star that drives the feedback and on the surrounding region: the surrounding interstellar medium (ISM) plays a crucial role in the physics and the detectability of the structures mentioned above. For example, after the massive star's supernova, accretion onto the remaining compact object is known to enact strong feedback on the surrounding ISM via high-energy radiation and outflows \citep[e.g.,][]{Munoz2016optical}. However, direct detections of such impact sites are rare and highly dependent on suitable ISM conditions \citep[particularly density and temperature;][]{2005Natur.436..819G,brown2005,2025A&A...696A.223A}. 

The most commonly observed feedback sites of massive stars, prior to their supernova explosions, are bow shocks: approximately $\sim 10^3$ massive star bow shocks and candidates are known from infrared imaging of the Galactic Plane \citep{peri2012,peri2015,kobulnicky2016,mw_project}, as well as a smaller number in the Small and Large Magellanic Clouds \citep{2010A&A...519A..33G,gvaramadze2011}. The formation process of bow shocks can explain why they are relatively easily detected. A bow shock forms when the stellar wind of a massive star collides with the star's surrounding ISM, in the presence of a relative, supersonic velocity difference between star and ISM \citep{1988ApJ...329L..93V,wilkin1996}. Importantly, the bow shock forms at the location where the stellar wind and ISM ram pressures are in balance. Ram pressure scales with density times velocity squared, and the stellar wind velocity typically exceeds the velocity difference between the star and the ISM by one to two orders of magnitude. Therefore, to compensate, the density at the bow shock site is dominated by ISM gas, which enhances the emissivity of the feedback site compared to stationary cases \citep{comeron1998,brown2005,vandeneijnden2022_racs}. Without the supersonic velocity difference with the ISM, the large (also supersonic) velocities of the stellar wind still create roughly symmetric shocks around the star \citep[see, e.g., the WR star G2.4+1.4;][]{Prajapati2019}. However, the significantly lower density in those wind shocks makes them hard to detect. 

Bow shocks around massive stars can be used for geometric inferences on the mass loss rate and velocity of the stellar wind, by equating the wind and ISM ram pressures \citep{wilkin1996,2019AJ....158...73K}. While relatively under-used \citep{vink2022}, this method can be highly complementary to methods that rely on modeling the wind spectra in bands from UV to radio. Bow shocks are also sites of particle acceleration, with radio observations in recent years revealing an increasing number of radio bow shocks and growing evidence for relativistic populations of electrons \citep{benaglia2010,moutzouri2022,vandeneijnden2022_velx1,2024MNRAS.532.2920V,2025A&A...704A.268M}. 

Moreover, the detection of a bow shock is routinely used to deduce the peculiar motion of the massive star that drives it: assuming the shock is aligned with the stellar motion, it reveals the direction and the supersonic nature of the peculiar motion. Therefore, bow shocks are often used to infer the runaway nature of a star: since the velocity required to escape the natal clusters of massive stars -- either via dynamical interactions \citep{poveda1967}, supernova kicks \citep{blaauw1961}, or a combination of both \citep[e.g.,][]{2026A&A...705A.215C} -- is similar to the speed of sound of the warm ISM ($\sim 10$ km s$^{-1}$), runaway stars may be expected to create a detectable bow shock in suitably dense ISM regions. In environments where the sound speed is significantly higher, e.g. in the hot ISM ($\sim 100$ km s$^{-1}$), a bow shock may not form \citep{2002A&A...383..999H}. At the other extreme, at high densities in ultracompact HII regions, the sound speed is much lower ($\sim 1$ km s$^{-1}$), and motion of the massive star embedded in the ultracompact HII region has been invoked to explain the cometary shape often observed in these structures \citep{1989ApJS...69..831W}.

The picture painted above, however, is only complete if the star's peculiar motion is the only, or dominant, driver of the bow shock. Observed bow shocks turn out to show more complex behavior: because \textit{relative} supersonic velocity drives the formation of bow shocks, they can also form \textit{in-situ}. In-situ bow shocks form when the star is not moving compared to its surrounding stars and other objects (i.e., HII regions), but instead a bulk ISM motion leads to the required supersonic net velocity \citep{2008ApJ...689..242P}. Because the low-density ISM itself is usually not detectable in the same band as the bow shock -- let alone that its bulk motion on parsec-scales can be observed -- this alternative scenario is challenging to identify. In addition to in-situ bow shocks, hybrid cases are expected to exist, where both ISM and massive star move compared to their local standard of rest. The dichotomy between \textit{in-situ} and \textit{runaway} bow shocks is, after all, merely a matter of definition of the two extreme combinations of ISM and stellar velocity \citep[see also][who recently argued for a dominant role of the ISM in creating bow shocks]{2025A&A...698A..64R}. 

In selected individual cases, the dominant role of ISM motion in creating bow shocks can be obvious. One example is the scenario of runaway stars escaping a stellar cluster at tens of km s$^{-1}$, while the stellar cluster also drives a cluster-scale wind at velocities of at least hundreds of km s$^{-1}$ \citep{2020Ap&SS.365....6K}. This combination causes a bow shock in the opposite direction of stellar movement as the cluster wind catches up with the star. An example of this effect was recently observed around the young massive cluster Westerlund 1, where \textit{JWST} imaging reveals bow shocks oriented towards the cluster, as well as direct evidence for a cluster-scale outflow \citep{2025A&A...693A.120G}. Similar examples can be found in distinctly different systems, such as the case of a young-stellar object with a bow shock caused by outflows from the Galactic Center (\citealt{2023ApJ...944..231P}; notably, this is a region where massive stars with bow shocks are also found orbiting the Galactic Center, e.g., \citealt{2014A&A...567A..21S}, \citealt{2019arXiv190300466Z}). 

In less extreme cases, the effect of ISM motions on the bow shock's morphology will be more subtle but still substantial \citep{1979ApJ...230..782G,2008ApJ...689..242P}. The key effect here is the change in net velocity difference between star and ISM: ISM motion causes a change in both the magnitude and orientation of this relative velocity field. As a result, the bow shock becomes misaligned from the direction of stellar peculiar motion. Furthermore, it alters the stand-off distance \citep{wilkin1996} where the stellar wind and ISM ram pressure are equal. Therefore, ISM motion will affect geometrical mass loss rate estimates, as well as conclusions about the kinematics and runaway nature of the massive star. Particle acceleration modeling also depends explicitly on the net velocity difference between the ISM and the massive star \citep{2024MNRAS.532.2920V}. 

Several previous works have investigated to what degree bow shocks align with the direction of stellar peculiar motion. Early examples use \textit{Hipparcos} astrometry for runaway stars or bow shocks up to relatively small distances, either with \citep[e.g.,][]{2018A&A...618A.110B} or without \citep{1995AJ....110.2914V,kobulnicky2016} explicit corrections for Galactic rotation. As the \textit{Hipparcos} astrometry suffers from relatively large uncertainties in proper motion, similar studies using \textit{Gaia} have been able to study larger samples to further distances \citep[see, e.g.,][]{2022AJ....164...86K,2025A&A...698A..64R,2026A&A...709A..97A}. The improved \textit{Gaia} accuracy and larger distances, importantly, necessitate a careful correction for Galactic rotation. In addition to these sample studies,  several works provide highly in-depth discussions of individual objects and their (misaligned) bow shocks \citep[see for instance,][]{2008ApJ...689..242P,2025A&A...698A..64R}. This suite of existing studies, in other words, spans different approaches to study bow shock alignment -- in terms of Galactic rotation correction, sample selection, sample size, and astrometric measurements. However, there is no unified framework that currently models the observed bow shock orientations across the entire population, combining both stellar and ISM motion. 

In this work, we propose such a framework in two steps: we first combine the two largest existing infrared bow shock catalogs \citep{kobulnicky2016,mw_project} with \textit{Gaia} DR3 astrometry and a new Galactic rotation correction, to measure peculiar velocities and bow shock misalignment across the sample (Section \ref{fig:data}). We then develop a statistical method to derive the ISM kinematics required to explain these observed bow shock orientations (Section \ref{sec:model}). We corroborate earlier results that only a small subset of bow shocks ($21\pm2\%$) fits the classic picture of a supersonic star with an aligned shock. Larger fractions of bow shocks would either not exist without ISM motion ($48\pm2\%$) or are greatly affected by it ($30\pm2\%$).  We also conclude that modest and realistic ISM motions at velocities of the order $\sim 10-15$ km s$^{-1}$ are sufficient to ensure that the ISM plays such a dominant role in bow shock formation and geometry.  Finally, we demonstrate how these motions affect mass-loss estimates of massive stars, specifically leading to underestimated rates at sub-sonic peculiar stellar velocity. 

\section{Observations: (mis)alignment angles and peculiar velocities in \textit{Gaia} DR3}

\label{sec:data}

In this paper, we investigate the orientations of bow shocks and the kinematics of the massive stars that drive them. In this Section, we present measurements of these two quantities: the peculiar velocity $v_{\rm star}$ of the star, and the angle $\beta$ between the direction of its peculiar motion and the apex of its bow shock. We discuss the selection of the studied bow shocks and stars (Section \ref{sec:target_selection}), introduce an improved correction for Galactic rotation (Section \ref{sec:astrometry}), and present the results of our analysis (Section \ref{sec:dataresults}).

\subsection{Target selection}
\label{sec:target_selection}

Bow shock orientations are measured from their images, most typically in infrared survey data. For our analysis, we use the bow shock samples of \citet{kobulnicky2016} and the Milky Way project \citep{mw_project}. These two papers present the largest catalogs of Galactic massive star bow shocks and candidates and report the orientation of the bow shock, defined as the direction of the apex in degrees East of North. As the typical uncertainty of these orientations we assume $5\degree$. While other papers report bow shock orientations or misalignment angles, they typically do so for comparatively small numbers of sources or using astrometry from \textit{Hipparcos} \citep{2007A&A...474..653V}. We instead use \textit{Gaia} DR3 \citep{2023A&A...674A...1G} and an updated method to measure the peculiar stellar motion compared to these papers (see Section \ref{sec:astrometry}); therefore, we build upon the two catalogs mentioned above. We stress that both catalogs were constructed from IR images and not by targeting runaway massive stars, introducing a selection effect that will be important in the Discussion of this work. 

We first select all 709 bow shocks from the \citet{kobulnicky2016} catalog and then add the 165 non-overlapping bow shocks from \citet{mw_project} with `Reliability Flag' equal to `R' (indicating the more-reliable subset). For the resulting sample of 874 bow shocks, we find the \textit{Gaia} object closest to the position of the bow-shock-driving star. We then select only the sources with a Gaia counterpart, e.g., with a separation less than $1$ arcsecond. We then apply \textit{Gaia} data quality cuts to the sample, following the approach in \citet{2023A&A...670A.108S} and \citet{2024A&A...681A..21S}. Importantly, we only select \textit{Gaia} counterparts with renormalised unit weight error (\textsc{ruwe}) less then $1.4$ to remove sources with poor astrometric solutions, including potential binaries. We also remove sources where the ratio of parallax over its error is smaller than five. After all selection steps, we are left with a sample of 233 bow shocks for our analysis. 

Our sample of 233 bow shocks is larger than the 139 bow shocks for which \citet{kobulnicky2016} investigate the distribution of the mis-alignment angle $\beta$, where only sources with proper motion uncertainties less than $45\degree$ are selected. In this \textit{Hipparcos}-based analysis, proper motion uncertainty dominates over the uncertainty of the bow shock orientation ($\sim 5\degree$). In our \textit{Gaia} analysis, the measured proper motions have negligible uncertainties in comparison. More recently, \citet{2022AJ....164...86K} investigated the orientation of bow shocks using the same two bow shock samples as we do, using \textit{Gaia} as well. We confirmed that our Gaia counterpart identifications are the same as the identifications in \citet{2022AJ....164...86K}. However, as we employ a new correction of Galactic rotation (introduced in the next Section), we do not use their peculiar motions or misalignment angles. 

\begin{figure*}
\includegraphics[width=\textwidth]{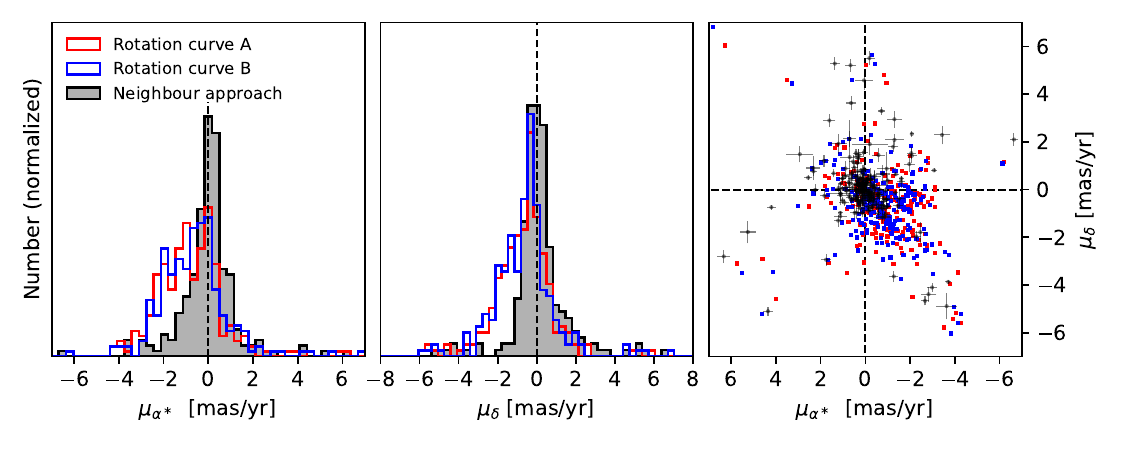}
 \caption{The peculiar angular motion of the considered sample of bow-shock-driving stars. We plot histograms of the motion in right ascension ($\mu_{\alpha*}$; left) and declination ($\mu_\delta$; middle), as well as both plotted against each other (right). The red and blue curves and points show the values corrected for Galactic rotation using a rotation curve model (e.g., Equation \ref{eq:corr}). The black curves and points have instead been corrected using the close-neighbor method.}
\label{fig:method}
\end{figure*}

\subsection{Astrometry: correcting for Galactic rotation}
\label{sec:astrometry}

The measured \textit{Gaia} proper motions  need correcting for their local Galactic rotation and the peculiar Solar motion. In previous analyses of bow shock orientations, particularly pre-\textit{Gaia}, these corrections were often not performed. As noted by \citet{1995AJ....110.2914V} -- the first paper to calculate the misalignment of bow shocks -- these corrections are small for massive stars within a few kpc and the uncertainties are dominated by measurement errors on the \textit{Hipparcos} proper motions. \citet{kobulnicky2016} similarly investigate the misalignment of bow shocks using uncorrected \textit{Hipparcos} proper motions. For our larger sample, which extends up to larger distances and has smaller \textit{Gaia} uncertainties on parallax and proper motion, this Galactic rotation correction becomes necessary. \citet{2022AJ....164...86K}, \citet{2024MNRAS.532.2920V}, \citet{2025A&A...694A.250C}, and \citet{2025A&A...698A..64R} are other examples where similar corrections are performed using \textit{Gaia} astrometry. 

Across those recent \textit{Gaia}-based studies, the commonly applied approach is to use a Galactic rotation curve model to perform this correction. Following Equations 2a and 2b from \citet{2007A&A...467L..23C}, who quote from \citet{1987pgim.book.....S}, the correction in Galactic coordinates is given by
\begin{equation}
\begin{split}
    \mu_{l*,\rm corr} = \text{ } 0.211 & \Bigg[A \cos 2l \cos b + B \cos b + {U \over D} \sin l - {V \over D} \cos l\Bigg] \\
    \mu_{b,\rm corr} = \text{ } 0.211 & \Bigg[-A \sin 2l \sin b \cos b + {U \over D} \cos l \sin b \\
    & + {V \over D} \sin l \sin b - {W \over
D} \cos b\Bigg] \text{ .}
\label{eq:corr}
\end{split}
\end{equation}
$A$ and $B$ are the Oort constants in km s$^{-1}$/kpc, $(U,V,W)$ describe the Solar peculiar motion in km s$^{-1}$ (radial, azimuthal, toward the Galactic North pole), $D$ is the source distance in kpc, and the numerical factor $0.211$ ensures that $\mu_{l*,\rm corr}$ and $\mu_{b,\rm corr}$ are calculated in mas/yr. The observed proper motion $\mu_{l*},\mu_{b}$ should be corrected by subtracting the values from Equation \ref{eq:corr}. Note that here, and throughout the remainder of this work, we use the following notations for brevity: $\mu_{l*} \equiv \mu_{l}\cos b$ and $\mu_{\alpha*} \equiv \mu_{\alpha}\cos \delta$.

\begin{figure*}
\includegraphics[width=\textwidth]{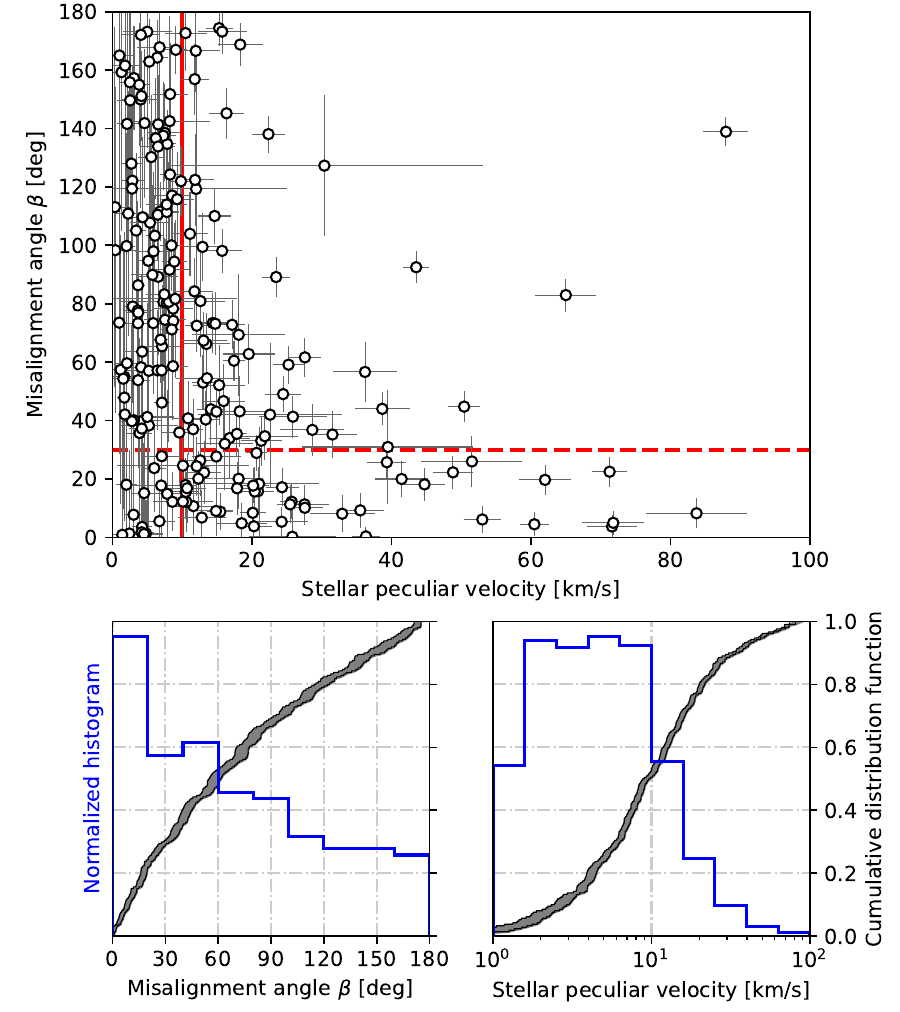}
 \caption{The main observational result of this paper: the measured peculiar velocities and bow shock misalignment angles for the 210 objects in our final sample. The top panel plots both quantities against each other, highlighting four quadrants with the red line (at $v_{\rm star} = 10 \text{ km s$^{-1}$ } \approx v_{\rm cs}$) and dashed line (at $\beta = 30\degree$). The bottom-left and bottom-right panels show the normalized histograms (blue) and cumulative distribution functions (CDFs; black) of $\beta$ and peculiar velocity, respectively. In the CDFs, the shaded region indicates the uncertainty, calculated as described in the main text.}
\label{fig:data}
\end{figure*}

For our sample of $233$ sources, we apply this rotation model correction using two sets of parameters: firstly (A), we take $(U,V,W) = (11.1,12.24,7.25)$ km s$^{-1}$ from on \citet{2010MNRAS.403.1829S} and $A=-B=12.5$ km s$^{-1}$ kpc$^{-1}$ \citep{2007A&A...467L..23C}. Secondly (B), we attempt $(U,V,W) = (10.6,10.7,7.6)$ km s$^{-1}$ and $A=15.7, B=-13.8$ km s$^{-1}$ kpc$^{-1}$, based on \citet{2025A&A...694A.250C}. For each source, we first take their proper motion in the Equatorial frame, as well as their Equatorial coordinates and \textit{Gaia} distance, to calculate their Galactic coordinates and proper motion in the Galactic frame. We then correct their proper motion using the two sets of rotation curve parameters above, and convert back to the Equatorial frame. We note that this paper is accompanied by a public repository where these steps can be repeated for clarity and reproducibility. 

In Figure \ref{fig:method}, we show the resulting peculiar angular motion in Equatorial coordinates $(\mu_{\alpha*}, \mu_\delta)$ as histograms and plotted against each other. The red and blue curves indicate the two models for Galactic rotation we attempted. It is notable that in both cases, the average $\mu_{\alpha*}$ and  $\mu_\delta$ are negative. As is visible in the right-hand panel, this would introduce a bias towards peculiar motion directions around $\sim 225 \degree$ East of North, e.g., towards the bottom right. Such a preferential direction within the Galaxy is not expected and indeed results from structured residuals in the Galactic rotation correction: the majority of our sources is located in the 1/3$^{\rm rd}$ of the Milky Way with $l \sim 200 - 320 \degree$. In that range, both of the attempted Galactic rotation models systematically under-predict the observed proper motion $\mu_{l*}$, leading to a non-zero mean in both $\mu_{\alpha*}$ and $\mu_\delta$ after changing coordinate frames. 

To improve beyond this bias in the peculiar motion, we instead employ a close-neighbor approach to calculate the Galactic rotation corrections. For each source, we select its closest neighbors in physical (3D) distance in the \textit{Gaia} database, considering only objects that match the same data quality selections as we applied to the bow shock driving stars. We initially select all neighbors within 20 parsec; if we find fewer than 50 neighbors, we increase the search radius with 20 parsec and repeat, up to a maximum of 100 parsec to ensure we sample a coherent part of the Galactic rotation curve. With this approach, we retain 210 sources, because a small number of sources does not have more than 3 neighbors with sufficient \textit{Gaia} data quality within 100 parsec. For each of these 210 sources, we calculate the median $\mu_{\alpha*}$ and median $\mu_\delta$ of its neighbors and subtract these values from the proper motion of the source. Using the median instead of the mean prevents individual outliers, such as other runaway stars, from significantly biasing the correction. 

The black lines and points in Figure \ref{fig:method} show the peculiar motions calculated with the neighbor method. The width of the peculiar angular motion distributions (left, middle) is smaller and their averages are consistent with zero. The right-hand panel shows that with this new method, no preferential direction of motion remains. An important effect of this method, however, is the magnitude of the uncertainties: when calculating the median proper motion of a source's $N$ neighbors, we conservatively assign it an uncertainty $\sigma/\sqrt{N}$, where $\sigma$ is the standard deviation in the $N$ measurements of the proper motion in either right ascension or declination. These uncertainties are, on average, $\sim 0.2$ mas/yr in both right ascension and declination, and greatly exceed the statistical uncertainties on the proper motion measurements of \textit{Gaia}. 

In all further computations, we take the uncertainties on peculiar angular motion in right ascension and declination (e.g., separately) as the basis for Monte-Carlo error propagation. This approach is essential when calculating orientation uncertainties: orientations are defined from $0\degree$ to $360\degree$ East of North, so otherwise orientations are assigned unrealistically large uncertainties when close to these limits (the same issue would arise for misalignment angles $\beta$ close to $0\degree$ or $180\degree$). 

We finish the \textit{Gaia} analysis by calculating the peculiar motion direction and velocity for all 210 remaining sources. We also calculate their misalignment angles $\beta$ by comparing with the bow shock orientations from the original bow shock catalogs \citep{kobulnicky2016,mw_project}. Here, we combine the error from astrometry and the orientation measurements in quadrature. 

\subsection{Measurement results}

\label{sec:dataresults}

In Figure \ref{fig:data}, we show the main results of our analysis in the previous Section: the measured peculiar velocities and misalignment angles $\beta$ for all remaining 210 systems in our sample. In the top panel, we show these two quantities plotted against each other, highlighting four quadrants: we highlight the regions with peculiar velocity above and below $10$ km s$^{-1}$ -- the typical speed of sound in the ISM -- as well as those with $\beta$ above and below $30\degree$. As we will show with our modeling in Section \ref{sec:model}, such small misalignment angles require the ISM to move slower than the star, and/or in the opposite direction of the star. The bottom right quadrant harbors \textit{classical} runaway bow shocks, with supersonic velocities and a bow shock in the direction of stellar peculiar motion. The top left quadrant is the opposite: slow-moving stars whose large bow shock misalignment shows that the local motion of the ISM is the cause of the bow shock (`\textit{in-situ}' bow shocks). The bottom left quadrant appears to be an extension of this top left quadrant where the ISM motion forms a head wind by chance, as we will corroborate later in our discussion (e.g., Figure \ref{fig:results_per_v}). Finally, the top right quadrant indicates supersonic systems with large misalignment angles, requiring a significant bulk ISM motion. 

To quantify the above statements, we turn to the bottom two panels of Figure \ref{fig:data}. Here, we show normalized histograms (blue) and the cumulative distribution functions (CDFs; gray) of misalignment angle $\beta$ (left) and peculiar velocity (right). For the CDFs, the shaded areas show the 67\% confidence level. We calculate these from $10^3$ randomized sets of the observed stellar peculiar motions, varied assuming Gaussian uncertainties. For each randomized set, we re-calculate the CDF of $\beta$ and $v_{\rm star}$. From all CDFs, we then calculate the 67\% confidence level in the quantity on the horizontal axis, at each plotted level in cumulative probability. We stress that error bars in a cumulative distribution run in horizontal direction: when plotting with the same number of bins as data points, as we do here, the cumulative probability levels are the same between iterations, but its corresponding horizontal value (e.g., $\beta$ or $v_{\rm star}$) changes. We use this same approach in all CDFs plotted in the remainder of this paper, both when plotting data or model ISM calculations (where we will show $95\%$ confidence levels to be conservative).

We find that $70\pm2\%$ of all bow shocks are misaligned, where we take $\beta =  30\degree$ as the minimum misalignment angle for this classification. Out of all $20\degree$ bins in $\beta$, the $0-20\degree$ bin is the most common, but, compared to all others combined, greatly in the minority. In the right-hand panel, we also find that $48\pm2\%$ of the bow-shock driving stars has sub-sonic peculiar velocity. This approximate half of all stars in the sample should not form a shock without additional ISM movement. Because the supersonic half of the objects also shows evidence for ISM influence (e.g., $\beta > 30\degree$), we conclude that the majority of massive star bow shocks are caused by or strongly shaped by the motion of the ISM: only $21\pm2$\% of systems is a \textit{classic} aligned runaway bow shock. The remaining $79\pm2$\% of bow shocks is strongly affected or (partially) caused by ISM motion. We note that we tested for, but did not find, any suggestion that the misalignment angles or peculiar velocities are related to the height of the system above the Galactic plane, although the final sample is dominated by systems located in the plane. 

Another way to consider these statistics, is to focus on runaways specifically. As the literature contains many definitions of runaway stars, based on their peculiar velocity, we do not take a single value for this classification. However, the lower right panel indicates that only $23\pm2$\% of stars in our sample moves faster than $20$ km s$^{-1}$. It is already well-known that the majority of runaway stars does not have an observed bow shock \citep{2002A&A...383..999H,peri2015,2025A&A...694A.250C}. Our analysis confirms, with new Galactic rotation corrections, that the opposite is true as well: the majority of massive star bow shocks is not caused by runaways. Bow shocks and runaways form a Venn diagram with little overlap, with the ISM as the main driver on both sides: ISM properties, such as low density or high temperature, can lead to undetectable bow shocks, while at the same time ISM motion can create bow shocks around stationary stars.

\section{Modelling the observed (mis)alignment of massive star bow shocks }
\label{sec:model}

\subsection{Rationale: what is the effect of bulk ISM motion on bow shock orientation?}
\label{sec:methodology}

\begin{figure}
\includegraphics[width=\columnwidth]{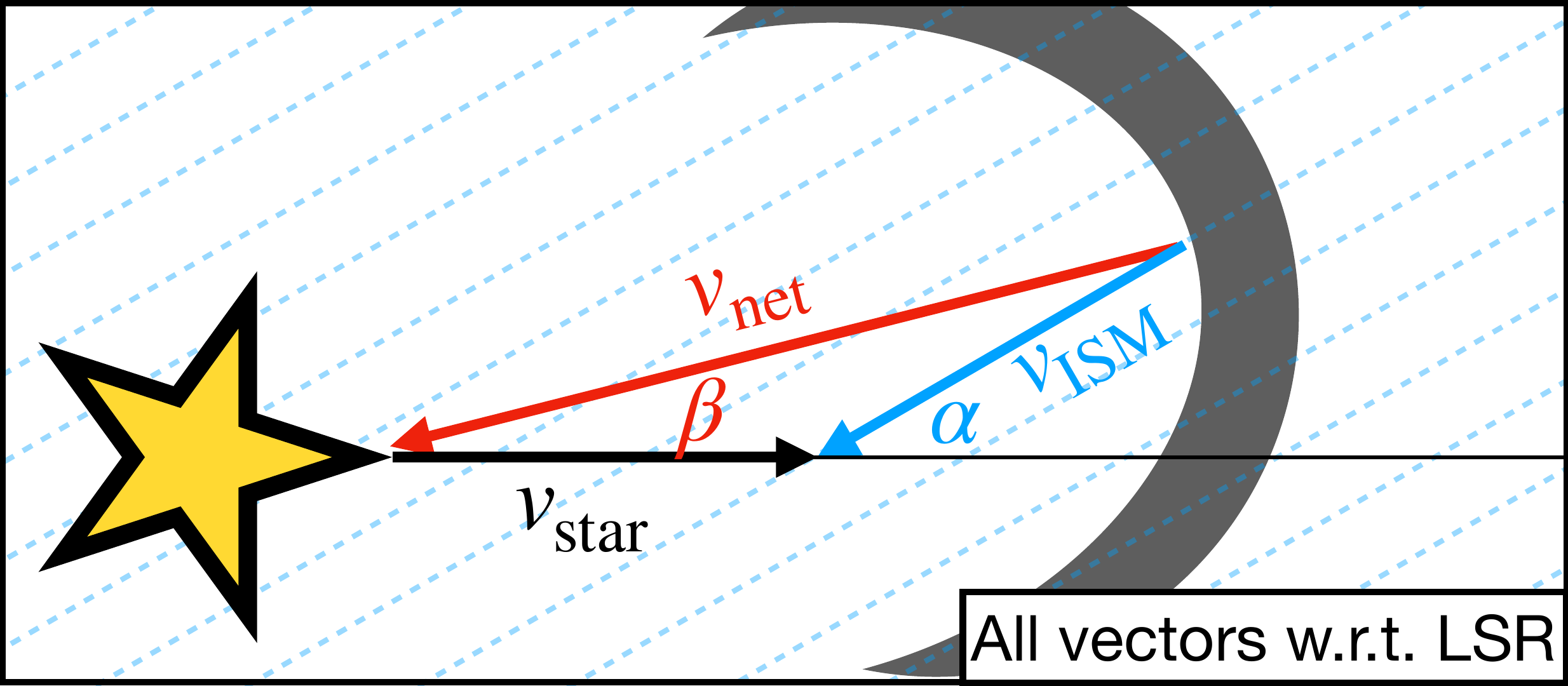}
 \caption{A schematic sketch of the geometry assumed in this work. The massive star moves at a peculiar velocity $v_{\rm star}$, in a direction that defines the zero-point of the bow shock misalignment. The ISM moves at a velocity $v_{\rm ISM}$ and an angle $\alpha$ relative to the direction of motion of the star. $\alpha = 0$ is defined as a head wind, leading to zero misalignment but decreasing the stand-off distance. We treat the problem as symmetric, e.g., considering $\alpha$ in the range $0-180\degree$. $\beta$ is the misalignment angle, calculated between the \textit{effective} direction of the incoming ISM and the stellar motion. Note that all these quantities are calculated \textit{in the LSR at the star's position}.} 
\label{fig:schematic}
\end{figure}

\begin{figure*}
\includegraphics[width=\textwidth]{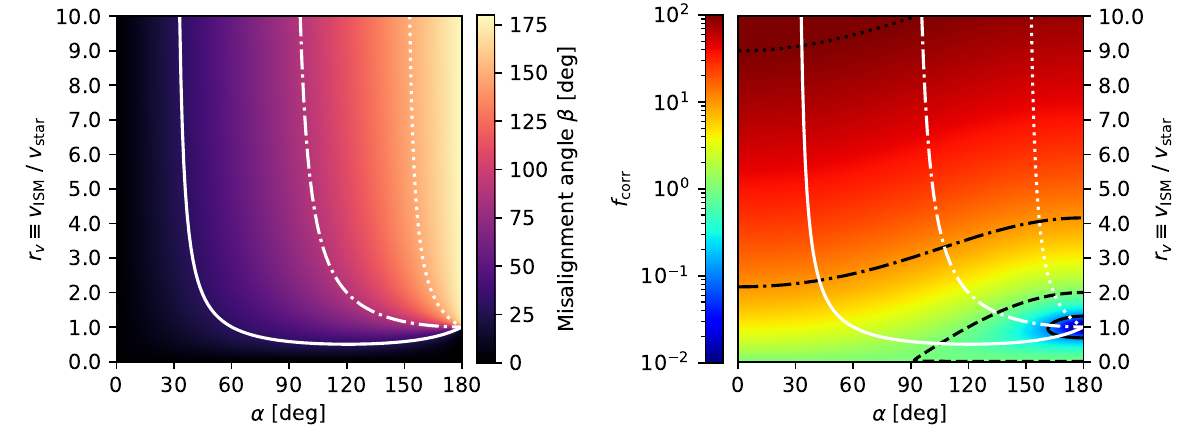}
 \caption{Colormaps of the misalignment angle $\beta$ (Equation \ref{eq:cosbeta}; left) and the correction factor $f_{\rm corr}$ to the stellar wind momentum flux (Equation \ref{eq:fcorr}; right). We plot both quantities for the same parameter space in relative ISM motion, defined by $r_v \equiv v_{\rm ISM}/v_{\rm star}$ and $\alpha$ (see Figure \ref{fig:schematic}). The white contours in the left-hand panel indicate, from left to right, $\beta = 30,90,150\degree$ and are copied in the right-hand panel. The black contours in the right-hand panel indicate $f_{\rm corr} = 1, 10, 100$. The classical picture of a runaway bow shock corresponds to the bottom left of both panels: a small $\alpha$ and $r_v \ll 1$. In that case, as expected, the misalignment angle $\beta$ is close to zero and the wind correction factor is close to unity.}
\label{fig:equations}
\end{figure*}

\begin{figure*}
\includegraphics[width=\textwidth]{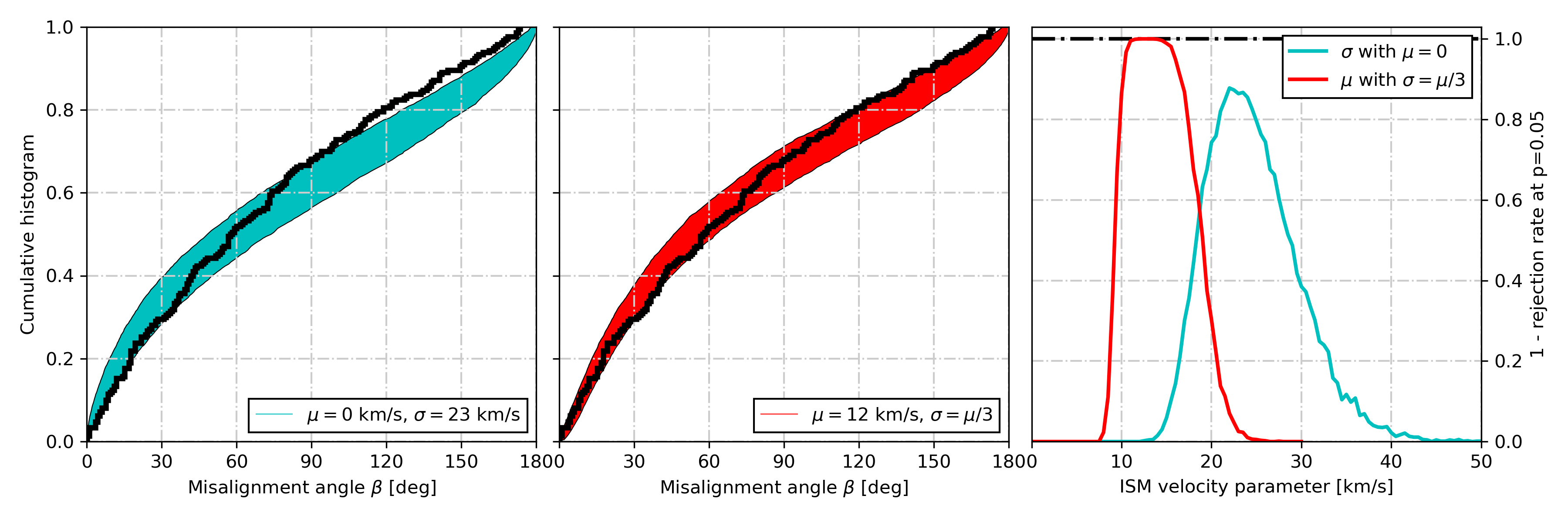}
 \caption{The result of Monte-Carlo simulations of the ISM movement. In the left-hand and middle panel, we show the cumulative distribution function of the observed misalignment angles $\beta$ as the black line. In both panels, the cyan and red regions show CDFs of simulated ISM cases. The left-hand panel assumes an ISM with a zero-centered Gaussian velocity profile, while the middle panel assumes a zero-truncated Gaussian velocity profile with width $\sigma = \mu/3$. We plot the simulated CDFs for $\sigma=23$ km s$^{-1}$ and $\mu=12$ km s$^{-1}$, corresponding to the best-matching ISM velocity profiles. The right-hand panel shows 1 minus the rejection rate of simulations as a function of $\sigma$ and $\mu$, indicating the level of $1$ with the dash-dotted line.}
\label{fig:results_all}
\end{figure*}

In this Section, we will discuss how bulk motion in the ISM affects the geometry and orientation of bow shocks around massive stars. We will focus our analysis on the (mis)alignment between the stellar motion and the bow shock apex, and on the stand-off distance between the star and the shock. Morphological bow shock properties beyond orientation and stand-off distance, such as its shape \citep[e.g., $R(\theta)$ as defined by][]{wilkin1996}, are not commonly measured for large samples of bow shocks \citep[although see][for a recent study into the asymmetries in bow shocks]{2025AJ....169..337W}. Such properties may highlight asymmetries in the ISM, such as its density (or velocity field), or in the stellar wind \citep{2000ApJ...532..400W}\footnote{We note that \citet{2000ApJ...532..400W} does not propose that asymmetries in bow shock shapes arise due to uniform, bulk ISM motion (as we discuss in our work), despite occasional mentions in later literature that it does.}. By focusing on misalignment and stand-off distance, we restrict our analysis to the assumption of symmetry: a single bulk motion of the ISM and a spherically symmetric stellar outflow\footnote{Some bow shocks may be caused by the pressure balance between the star's radiation field and the ISM ram pressure \citep{1988ApJ...329L..93V,2014A&A...563A..65O,2014A&A...566A..75O,2019MNRAS.489.2142H}. In particular, for late O-type stars with relatively weak winds, this radiation pressure may exceed the stellar wind's ram pressure. While this case affects the discussions regarding $R_0$ and $f_{\rm corr}$ in our work, a dominant radiation field does not violate our assumption of spherical symmetry.}. 

Before introducing our modeling approach, we note that we treat the effect of the ISM motion in two dimensions, e.g., in the plane of the sky. In the absence of radial velocities and bow shock inclination measurements for the majority of systems, we deem this the most reliable approach. Furthermore, it allows us to limit the problem to two parameters, as shown below: the ISM bulk velocity and angle with respect to the stellar motion. This simplication comes at the cost of potentially underestimating ISM velocities by a small factor. We note, however, that low inclinations (e.g., substantial radial motions) lead to deviations from the typical arc-like shape \citep{2016MNRAS.456..136A,2018MNRAS.477.2431T}, which was used to find the studied bow shocks (see Section \ref{sec:data}; \citealt{kobulnicky2016}, \citealt{mw_project}). Therefore, we assume here that this effect is relatively minor. 

In our treatment of this problem, we work in the star's Local Standard of Rest (LSR). When discussing peculiar motion of the star, or bulk motion of the ISM, we refer to motion relative to this LSR. In the classical picture, the ISM is assumed to be stationary (e.g., moving with the Milky Way). Therefore, the problem is usually reversed from a star moving at $v_{\rm star}$ through the ISM to a stationary star experiencing an ISM head wind at $-v_{\rm star}$. The balance of ram pressures of the ISM and the spherical stellar wind defines the classical stand-off distance \citep[e.g.][]{1971SPhD...15..791B,wilkin1996,delvalle2012}: 
\begin{equation}
R_{0, \rm classical} = \sqrt{\frac{\dot{M}_{\rm wind} v_{\rm wind}}{4\pi\rho_{\rm ISM}v_{\rm star}^2}}\text{ ,}
\label{eq:R0class}
\end{equation}
where $\dot{M}_{\rm wind}$ and $v_{\rm wind}$ are the stellar wind mass loss rate and terminal velocity, respectively, while $\rho_{\rm ISM}$ is the ISM density. The apex of the bow shock forms in the direction of the stellar peculiar motion. Evidently, the bow shock only forms for supersonic stellar peculiar velocities, e.g., $v_{\rm star} \gtrsim 10$ km s$^{-1}$ for typical ISM conditions. 

In Figure \ref{fig:schematic}, we sketch the situation if the ISM shows bulk motion within the LSR at a velocity $v_{\rm ISM}$, at an angle $\alpha$ with respect to the peculiar motion of the star. We define $\alpha = 0$ as a head-wind, which would decrease the stand-off distance by effectively replacing $v_{\rm star}$ with $v_{\rm star} + v_{\rm ISM}$ in Equation \ref{eq:R0class}. By again transforming to the frame where the star is stationary, we can derive the misalignment angle $\beta$ between the apex of the bow shock and the direction of stellar peculiar motion. For this calculation, we calculate the net combined ISM velocity due to its bulk motion and the stellar motion, and make use of the spherical symmetry of the stellar wind. From trigonometry, the misalignment angle $\beta$ can be written as 
\begin{equation}
\cos \beta = \frac{1 + r_v \cos \alpha}{\sqrt{1 + r_v^2 + 2r_v\cos \alpha}} \text{ ,}
\label{eq:cosbeta}
\end{equation}
where we define the velocity ratio $r_v \equiv v_{\rm ISM}/v_{\rm star}$. In other words, $r_v \gg 1$ implies that the ISM bulk motion is the dominant factor in determining the bow shock properties. In this limit of large $r_v$, Equation \ref{eq:cosbeta} displays the expected asymptotic behavior of $\beta \rightarrow \alpha^{-}$ (e.g., $\beta$ approaches but never exceeds $\alpha$): for very fast ISM motion, it fully dominates the direction $\beta$ of the apex, but never pushes it beyond the misalignment angle  $\alpha$  between ISM and stellar motions.

In the left panel of Figure \ref{fig:equations}, we plot a colormap of misalignment angle $\beta$ for the parameter space of $\alpha$ and $r_v$. The white contours indicate $\beta = 30\degree,90\degree,150\degree$ and clearly highlight the asymptotic behavior mentioned above. Notably, a measured misalignment angle $\beta \geq 90\degree$ requires that the ISM bulk motion dominates over the stellar motion, regardless of their relative angles. As expected, small misalignment is only seen for low ISM motion and/or small $\alpha$. For instance, as expected, a head-wind ISM ($\alpha \approx 0\degree$) moving at a fraction of the stellar velocity ($r_v \ll 1$) leads to a small misalignment angle $\beta$ (bottom left of the left panel in Figure \ref{fig:equations}). 

The bulk motion of the ISM will also change the ram pressure balance that defines the stand-off distance. By calculating the magnitude of the net combined ISM velocity, we can derive that 
\begin{equation}
R_0 = \frac{R_{0,\rm classical}}{\sqrt{1 + r_v^2 + 2r_v\cos \alpha}}\text{ ,}
\label{eq:R0new}
\end{equation}
where $R_{0,\rm classical}$, given by Equation \ref{eq:R0class}, is the stand-off distance for a stationary ISM and the stellar wind parameters. As expected, a large $r_v$ causes the stand-off distance to decrease with respect to the expected case, for given stellar wind properties. In this work, we will investigate a parameter we call $f_{\rm corr}$, which we define as 
\begin{equation}
f_{\rm corr} \equiv 1 + r_v^2 + 2r_v\cos \alpha \text{ .}
\label{eq:fcorr}
\end{equation}

\noindent In other words, $f_{\rm corr} \propto (R_{0,\rm classical} / R_0)^2$. By considering Equation \ref{eq:R0class}, we can conclude that $f_{\rm corr}$ equals the factor by which $\dot{M}_{\rm wind} \times v_{\rm wind}$ (the wind's momentum flux) is underestimated if bulk ISM motion is ignored: for a large $f_{\rm corr}$, the ISM motion is dominant, meaning that ignoring ISM motion leads to an underestimate of the stellar wind ram pressure. The full expected behavior of $f_{\rm corr}$ for the parameter space of $r_v$ and $\alpha$ is shown in the right-hand panel of Figure \ref{fig:equations}. The black contours show the $f_{\rm corr}=1,10,100$ levels. The combined stellar wind parameters $\dot{M}_{\rm wind} \times v_{\rm wind}$ are underestimated across most of the plotted parameter space, with the exception of tail-winds where $v_{\rm ISM} \approx v_{\rm star}$: in that scenario, the net ISM ram pressure experienced by the stellar wind decreases so the stand-off distance increases compared to the stationary ISM scenario. 

If the stellar wind parameters (e.g, $\dot{M}_{\rm wind} v_{\rm wind}$) and ISM particle density $n_{\rm ISM}$ are known from independent observations, it is possible to uniquely invert Equations \ref{eq:R0class}, \ref{eq:cosbeta}, and \ref{eq:R0new} to determine $\alpha$ and $v_{\rm ISM}$ from the observables $\beta$, $v_{\rm star}$ and $R_0$. Specifically, by combining $R_0$, the stellar wind parameters, and the ISM density, the total relative velocity $v_{\rm total}$ can be calculated using ram pressure balance. Trigonometry based on Figure \ref{fig:schematic} then shows that
\begin{equation}
    v_{\rm ISM} = \sqrt{v_{\rm total}^2 + v^2_{\rm star} - 2v_{\rm total}v_{\rm star}\cos\beta} \text{ .}
    \label{eq:vism}
\end{equation}
In this work, we do not focus on such calculations for individual systems, as knowing the stellar wind and ISM parameters accurately and independently is a major challenge\footnote{Equation \ref{eq:vism} deviates from the equation presented in Section 7 of \citet{2025A&A...698A..64R} for the same quantity. In our parameter definitions, this equation reads $v_{\rm ISM} = v_{\rm total} - v_{\rm star}\cos\beta$. Both versions converge towards each other for the cases of $\beta = 0$ and $\beta = \pi$. At other observed misalignment angles, our derivation predicts larger ISM bulk motion, especially for systems where the total velocity vector is dominated by the stellar peculiar velocity ($v_{\rm star}$ $\rightarrow$ $v_{\rm total}$).}.

\begin{figure*}
\includegraphics[width=\textwidth]{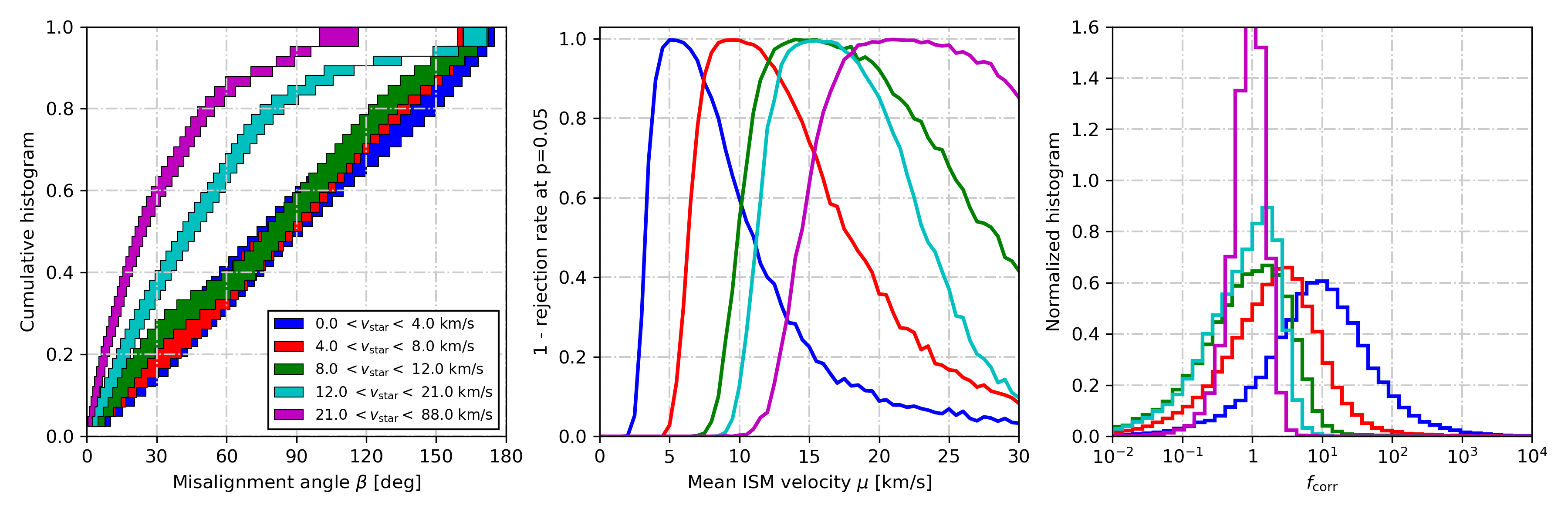}
 \caption{The analysis and ISM simulations per peculiar velocity quintile. The left-hand panel shows the observed CDF of $\beta$ per velocity quintile. The middle panel shows the acceptance rate for ISM simulations as a function of mean ISM velocity $\mu$, per quintile. The right-hand panel shows histograms of $f_{\rm corr}$ for every ISM simulation at $\mu = 12$ km s$^{-1}$, again per quintile.}
\label{fig:results_per_v}
\end{figure*}
    
\subsection{Methodology: how to constrain ISM motion from observables}

As the misalignment of bow shocks is caused by two ISM properties -- its speed and direction, both relative to the stellar peculiar motion in the LSR -- the measured misalignment angle $\beta$ cannot uniquely constrain the ISM motion. However, as we study a sample of 210 bow shocks, we can instead focus on their statistical properties: can we find a way to describe the average motion of the ISM in the LSR of each considered star, that can reproduce the observed \textit{distribution} of misalignment angles? To approach this question with the least number of assumptions, we postulate that the direction of the ISM motion ($\alpha$, defined with respect to the stellar peculiar motion) is distributed uniformly. The axisymmetry of the problem sets the upper limit of this uniform distribution at  $180\degree$. Its lower limit differs for each individual system: as the misalignment can never exceed the angle between ISM and stellar motion, $\alpha$ follows a unique uniform distribution between $[\beta, 180\degree]$ for each system. For the ISM velocity distribution we attempt two options: the positive half of a zero-centered Gaussian of width $\sigma$, and a zero-truncated Gaussian with mean $\mu$ and a width equal to a certain fraction of $\mu$. Here, we adopt $\sigma = \mu/3$.

We investigate what ISM properties can match the observed misalignment angles with the following steps. For a given $\sigma$ or $\mu$ of the velocity distribution in km s$^{-1}$, we draw 210 random ISM directions $\alpha$ and velocities $v_{\rm ISM}$. At the same time, for each of the 210 bow shocks driving stars, we draw a peculiar motion in RA and Dec from its measured value and uncertainties, and then calculate the corresponding velocity and misalignment angle (thereby taking the observational uncertainties into account in our analysis). The 210 simulated ISM motions are then combined with the 210 observed peculiar velocities to calculate what their misalignment angles would be. We then use a Kolmogorov-Smirnov test to assess whether the simulated and observed misalignment angles may arise from the same underlying distribution: we reject this hypothesis if we find $p<0.05$. By repeating this setup $1000$ times, we can calculate the rejection rate for a given ISM velocity distribution. We note that the $p=0.05$ limit is relatively high on purpose: for e.g., $p=0.01$ or smaller, we find that none of the 1000 iterations are rejected for a relatively broad range in $\sigma$ or $\mu$. Setting $p=0.05$ increases our ability to distinguish in the level to which the simulations reproduce the observed $\beta$ distributions. 

\subsection{Modelling results}

\subsubsection{Complete sample}
\label{sec:fullsamplefitting}

The results of our modelling analysis for the full sample are summarized in Figure \ref{fig:results_all}. These results highlight that a modest bulk motion in the ISM, with a mean velocity of $\mu \sim 12$ km s$^{-1}$, is sufficient to consistently explain the observed  distribution of misalignment angles. In the next two paragraphs, we explain further how we reach this conclusion. 

We performed the described analysis for a range of ISM velocity distributions: zero-centered Gaussians with $\sigma$ from $1$ to $50$ km s$^{-1}$, as well as zero-truncated Gaussians with $\mu$ between $1$ to $30$ km s$^{-1}$ and $\sigma = \mu/3$. In Figure \ref{fig:results_all}, we show the results of this analysis in three panels. In the left-hand and middle panels we show the Cumulative Distribution Function (CDF) of the observed misalignment angles $\beta$ in black. The cyan CDF in the left-hand panel shows the 95\% confidence level of the CDF for a simulated ISM with $\sigma = 23$ km s$^{-1}$. The middle panel shows the same result for an ISM with $\mu=12$ km s$^{-1}$ and $\sigma = \mu/3 = 4$ km s$^{-1}$. The right hand panel shows 1 minus the rejection fraction for both ISM velocity profiles, showing the peaks at $\sigma = 23$ km s$^{-1}$ and $\mu = 12$ km s$^{-1}$. 

From the right hand panel, two conclusions can be drawn: firstly, the plotted examples in the left-hand and middle panel are the examples that best match the observed distribution. Secondly, the zero-truncated Gaussian velocity profile (red) provides a systematically better match than a zero-centered Gaussian velocity profile (cyan). There exists no single $\sigma$ for which every simulation returns a satisfactory match to the observed CDF, while that is the case for a range in $\mu$. This difference is also reflected in the better match between observed and simulated CDF in the middle panel than in the left-hand panel. For this reason, we will continue our work using only the zero-truncated Gaussian with mean $\mu$ for the ISM velocity distribution.

\subsubsection{Results per velocity quintile}
\label{sec:per_v}

We continue our analysis by splitting the full target sample into five groups of equal size, based on their peculiar velocity. Specifically, we use peculiar velocity quintiles, which correspond to $<4$ km s$^{-1}$, $4-8$ km s$^{-1}$, $8-12$ km s$^{-1}$, $12-21$ km s$^{-1}$, and $>21$ km s$^{-1}$. The CDFs of their measured misalignment angles are shown in the left-hand panel of Figure \ref{fig:results_per_v}. It is evident that the three lowest-velocity quintiles follow similar distributions, close to a diagonal -- the CDF of a uniform distribution. This finding corroborates our earlier statement (Section \ref{sec:dataresults}) that for sub-sonic velocities, the aligned and misaligned bow shocks appear to form a continuation of the same distribution. In other words, the aligned bow shocks at sub-sonic velocity appear to arise due to chance alignment of the ISM and peculiar motion ($\alpha = 0$, i.e., a head wind). For the two supersonic velocity quintiles, the CDF looks remarkably different: a clear preference for smaller misalignment angles can be observed. In these two quintiles, the average $r_v$ for a given ISM velocity field, is smaller than in the three lower-velocity quintiles. As a result, misalignment of the bow shock is less common and the shock more likely reflects the presence of a classical runaway. 

We then repeat our investigation of the ISM motion for each velocity quintile. In the middle panel, we plot 1 minus the rejection rate of simulated ISM properties as function of the mean ISM velocity $\mu$. Unsurprisingly, we find that larger velocities of the ISM would be inferred for increasing quintiles in stellar peculiar velocity. While this finding has no physical implication, it highlights the importance of selecting our sample based on \textit{bow shock detection}. If, instead, we had considered only bow shocks around runaway stars, we would have inferred a significantly faster ISM velocity.  

We also turn to $f_{\rm corr}$: the factor by which we underestimate the stellar wind momentum flux ($\dot{M}_{\rm wind} v_{\rm wind}$) by ignoring ISM motions. For this purpose, we take $\mu = 12$ km s$^{-1}$ from Section \ref{sec:fullsamplefitting}. We again simulate 1000 iterations of the ISM motion per target and calculate the resulting $f_{\rm corr}$ for each, following Equation \ref{eq:fcorr}. We then plot histograms of $f_{\rm corr}$ per velocity quintile, with bins defined in logarithmic space, in the right-hand panel of Figure \ref{fig:results_per_v}. As expected, the typical under-estimate of stellar wind momentum flux is higher for stars at lower peculiar velocities. For those, $\dot{M}_{\rm wind} v_{\rm wind}$ is easily underestimated by more than an order of magnitude, although the distribution is broad. The mean correction factors and their $1\sigma$ confidence intervals are, in order of increasing velocity quintile: $\log f_{\rm corr} = 0.9^{+0.7}_{-0.2}, 0.3^{+0.6}_{-0.7}, -0.2^{+0.7}_{-0.6}, -0.2^{+0.5}_{-0.5}$, and $-0.1^{+0.3}_{-0.2}$. In general, this analysis shows that strong caution should be taken for any bow shock whose driving star does not have supersonic peculiar motion: only for the combination of high peculiar velocity and an aligned bow shock can we have confidence that $f_{\rm corr} \approx 1$. 

\subsubsection{Results per environment type}

\begin{figure}
\includegraphics[width=\columnwidth]{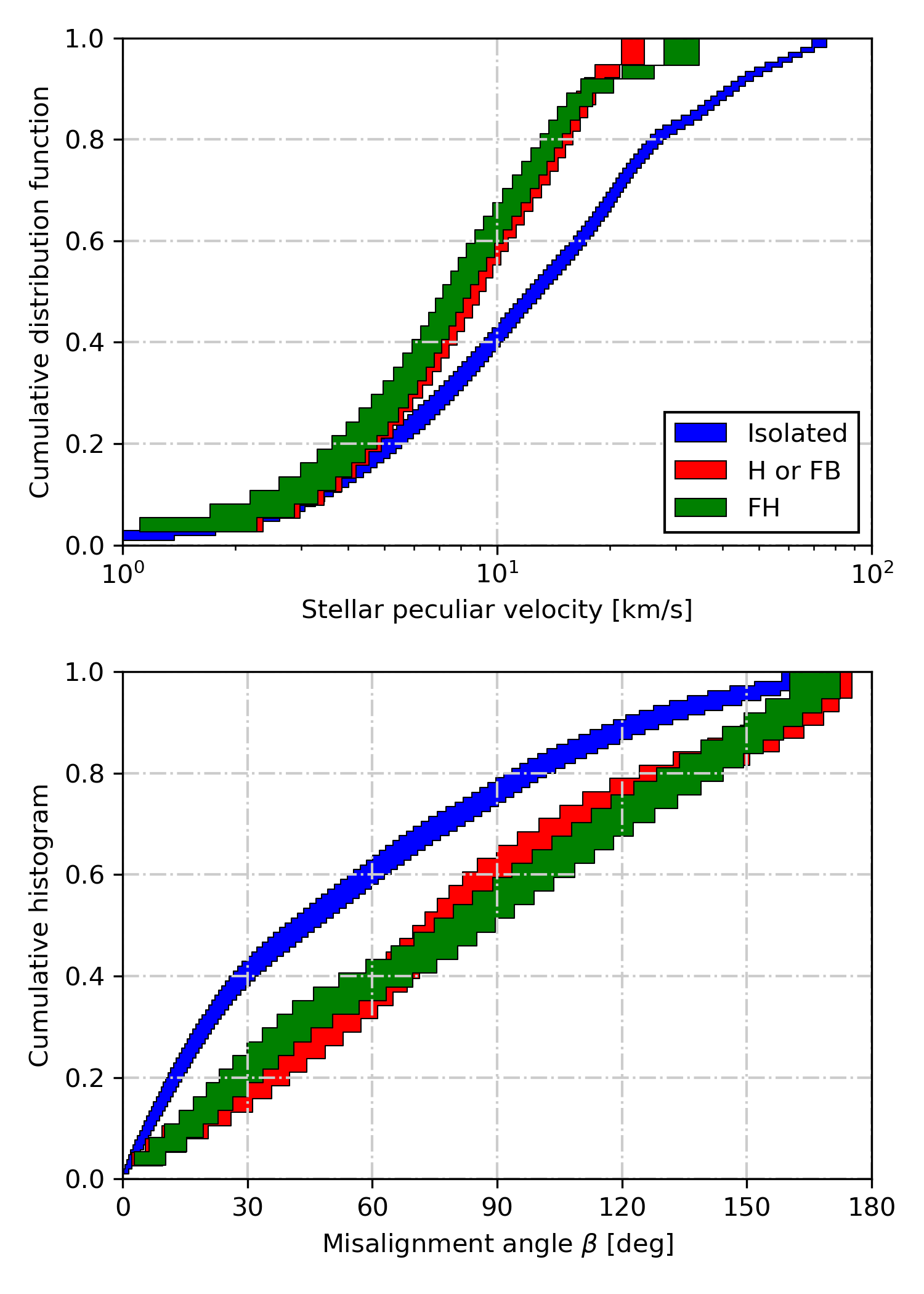}
 \caption{Cumulative distribution functions of the peculiar velocity (top) and misalignment angle $\beta$ (bottom), splitting the sample by environment class. Here, `I', in blue, indicates isolated systems; `H or FB', in red, indicates systems inside an HII region and/or facing a bright-rimmed cloud;  `FH', in green, indicates systems facing an HII region.}
\label{fig:v_per_env}
\end{figure}

\begin{figure*}
\includegraphics[width=\textwidth]{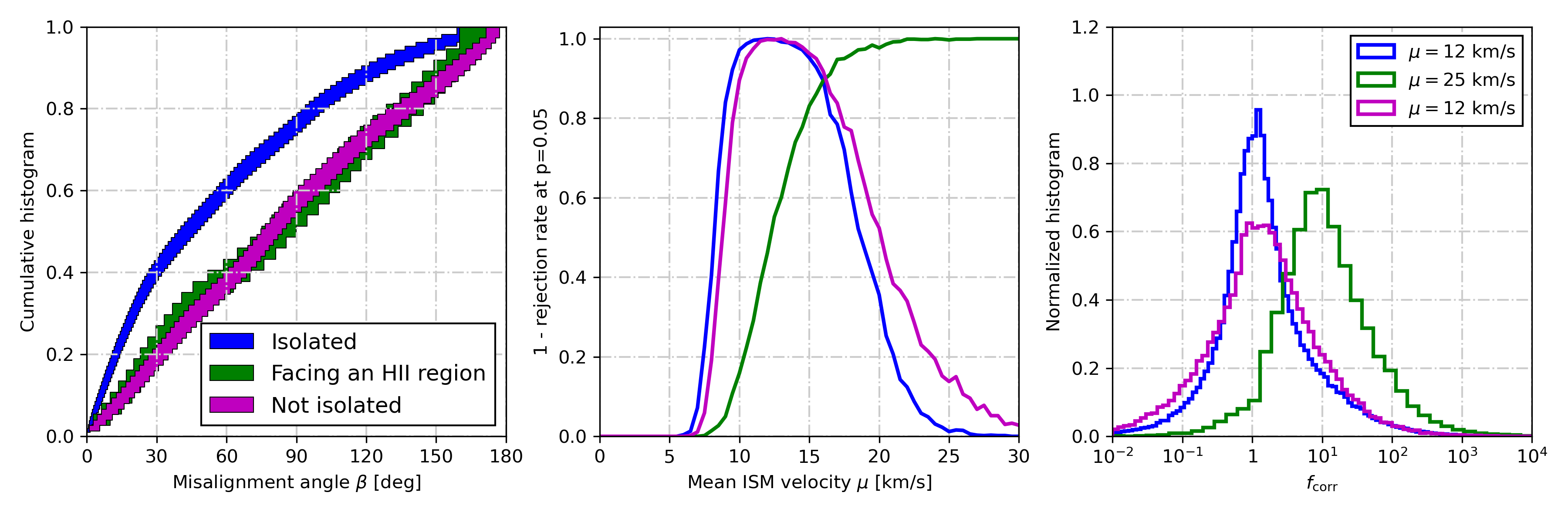}
 \caption{Same as Figure \ref{fig:results_per_v}, now splitting up the sample into Isolated (blue) and Non-isolated systems (cyan). In addition, in green, we show the result for systems facing an HII region, where we additionally assumed the ISM movement originates from the HII region (e.g., $\beta \approx \alpha)$.}
\label{fig:results_per_env}
\end{figure*}

Instead of splitting the sample into velocity quintiles, we can also split it up by type of environment. These environment types are derived by the original authors for the samples from \citet{kobulnicky2016} and \citet{mw_project}. We divide the sample into three classes from these environment types, in similar fashion to \citet{2022AJ....164...86K}: `I', for isolated systems; `H or FB' for systems inside an HII region and/or facing a bright-rimmed cloud (which may overlap in classification and are therefore combined); and `FH', for systems facing an HII region. This leaves 180 systems of our original sample, as the others have unknown environment classification or do not confidently fall into either category. It is pertinent to repeat the call for caution issued by \citet{2022AJ....164...86K} regarding these classifications: as they are based on \textit{angular} proximity to other infrared features, they may not always reflect the true local conditions of each bow shock. 

In the top panel of Figure \ref{fig:v_per_env}, we show the CDFs of our calculated peculiar velocities per environment type. We find a clear difference in the distributions of isolated systems (blue; Isolated) compared with the other two classes. Specifically, systems facing HII regions (green; FH) and either inside an HII region or facing a bright rimmed cloud (red; H or FB) have relatively lower peculiar velocities. Systems in the latter two environment types also appear consistent with each other in terms of their velocity distribution. This trend is not necessarily surprising, as larger peculiar velocities can cause systems to travel from their natal environment and become isolated. In the bottom panel of the same Figure, the CDFs of the misalignment angle $\beta$ reveal behavior similar to the peculiar velocity: the isolated systems have a different distribution than both classes with `structured' environments, while those latter two classes appear consistent with each other. The `FH' and `H or FB' also appear consistent with a uniform distribution in misalignment angle, as expected for systems dominated by low peculiar velocity. 

Due to the consistency in misalignment angles and peculiar velocities between the `FH' and `H or FB'  environment types, we combined these classes for the remaining analysis into a 'Non-isolated' class\footnote{We note that similar distributions in $v_{\rm star}$ and $\beta$, independently, could still correspond to a different distribution in the stellar pairs of $(v_{\rm star}, \beta)$. We confirmed that both classes separately returned fully consistently results when not combined in the further analysis.}. In Figure \ref{fig:results_per_env}, we repeat the analysis of the previous Section \ref{sec:per_v} for this `Non-isolated' versus the `Isolated' class (magenta and blue). Unsurprisingly, given Figure \ref{fig:v_per_env}, the isolated systems show more preference for aligned bow shocks than non-isolated systems. However, the middle panel shows that the bow shock misalignment of both types of systems can be explained by the same ISM motion. As shown in the right hand panel, this translates to similar distributions in $f_{\rm corr}$ as well. Because we have not split the sample by peculiar velocity in this case, the two $f_{\rm corr}$ distributions center at $1$ but spread across multiple orders of magnitude. 

In Figure \ref{fig:results_per_env}, in green, we also show the results for the FH class only, i.e., for systems facing an HII region. Here, we have made an additional assumption on the direction of the ISM: we assume that the ISM motion driving the bow shock orientation is due to an outflow from the HII region, which means we can use that $\alpha \approx \beta$. We can therefore repeat our analysis on the FH class without drawing random $\alpha$ but instead setting it equal to $\beta$. The results of this analysis are shown as the green curves in Figure \ref{fig:results_per_env}. Clearly, a larger velocity, $\mu \gtrsim 25$ km s$^{-1}$, is inferred for these HII region outflows. As a result, the wind momentum flux in these systems would be, on average, underestimated by an order of magnitude: $\log f_{\rm corr} = 1.0\pm0.6$. 

\section{Discussion}

In this paper, we quantitatively investigate what ISM motions can explain the observed misalignment of massive star bow shocks with respect to the stellar motion. To perform this analysis, we present a new approach to calculate the peculiar angular motion of the bow-shock driving stars, based on the observed proper motions of their close neighbors. We find that modest ISM movements of the order $\sim 10-15$ km s$^{-1}$ can reproduce the observed distribution of misalignment angles. This velocity also ensures that the relative velocity difference between the ISM and massive star is supersonic across the sample, even though the peculiar velocity of approximately half of the stars in our sample is sub-sonic. In this Discussion, we will first compare our results with earlier analyses and discuss differences in methodology, especially regarding the Galactic rotation correction. We will finally turn to the interpretation of the our results and their implications for ISM, massive star, and bow shock physics.

\subsection{Statistics of peculiar motion and bow shock formation}

\begin{figure}
\includegraphics[width=\columnwidth]{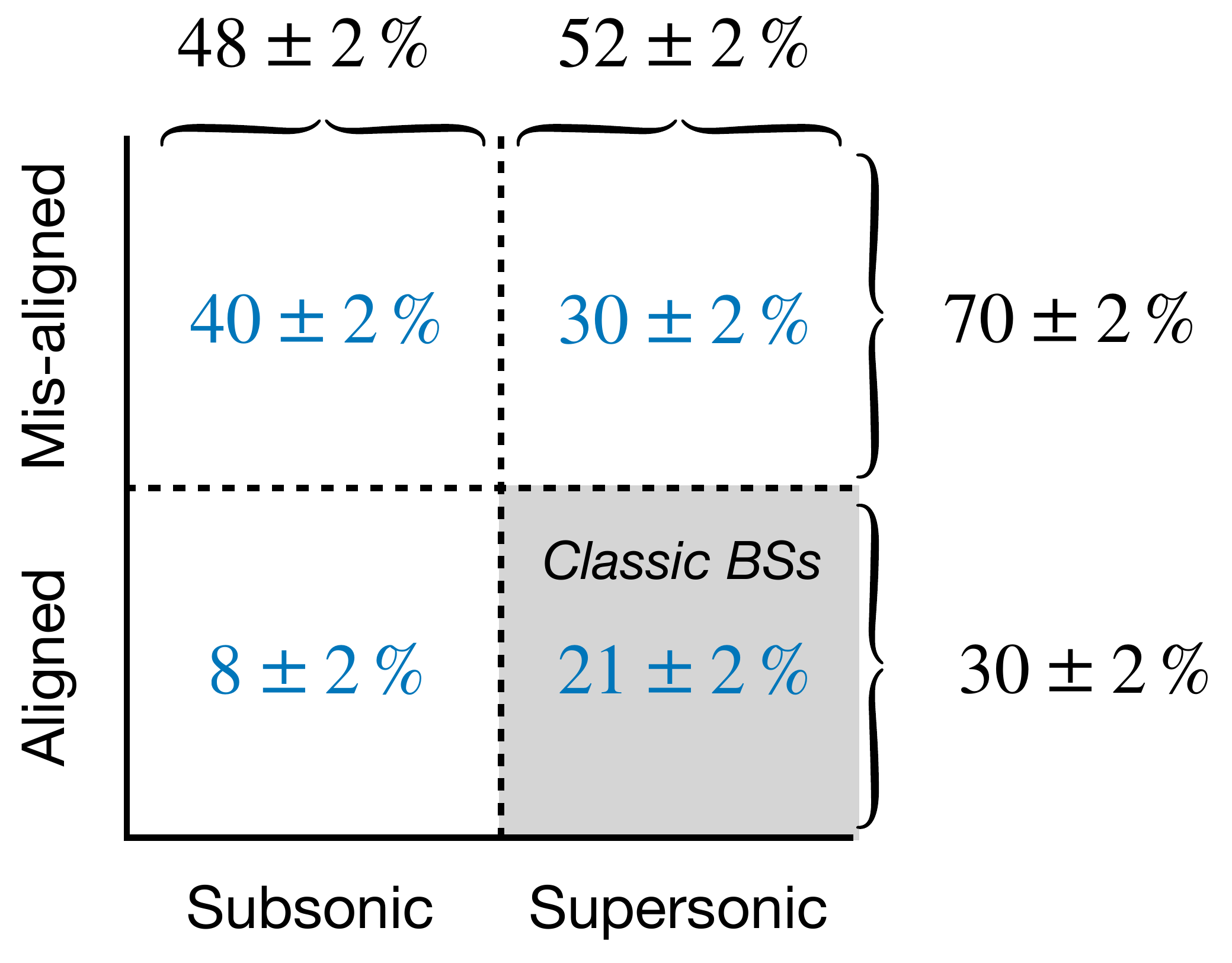}
 \caption{The basic statistics summarizing the peculiar velocities and mis-alignment angles across our sample. We take $10$ km s$^{-1}$ and $30\degree$ as the speed of sound and maximum angle for alignment, respectively. The percentages shown here are calculated from the Monte-Carlo error calculations underlying the histograms in the bottom panels of Figure \ref{fig:data}. Note that, due to rounding effects, not all numbers across rows or columns add up perfectly.}
\label{fig:schematic_perc}
\end{figure}

In terms of statistics, we find that only $23\pm2\%$ of the bow-shock driving stars in our sample would be classified as a runaway star when $20$ km s$^{-1}$ is taken as the threshold velocity. $48\pm2$\% of our sample moves at $v_{\rm star} \leq 10$ km s$^{-1}$. The angle between stellar motion and bow shock apex are misaligned (e.g., $\geq 30\degree$) for $70.1^{+1.8}_{-2.1}\%$ of stars in our sample. We summarize these basic statistics of our sample in Figure \ref{fig:schematic_perc}, for which we highlight that only $21\pm2\%$ of bow shocks in our system can be classified a classic bow shock, i.e., aligned and driven by supersonic peculiar motion. Another noteworthy point is that the ratio of aligned and mis-aligned bow shocks for sub-sonic massive stars, is consistent with one to five -- the same ratio as the range in angles, $30\degree$ versus $150\degree$, spanned in our definitions of aligned and mis-aligned shocks. In other words, for sub-sonic systems, the orientation of the bow shock is consistent with uniform, which further corroborates that the dominant factor in its creation is the ISM motion. 

Our statistics are broadly consistent with earlier studies\footnote{Note that we do not explicitly compare with studies based on \textit{Hipparcos} data and/or studies that do not correct for Galactic rotation, e.g., \citet{1995AJ....110.2914V}, \citet{kobulnicky2016}, or \citet{2018A&A...618A.110B}.} on the same or similar samples: \citet{2022AJ....164...86K}, for instance, report a runaway fraction of $24^{+9}_{-7}$\% taking a definition of $25$ km s$^{-1}$, therefore finding a slightly higher fraction than we do. \citet{2019AJ....158...73K} do not explicitly focus on runaway fraction, but show a peculiar velocity distribution peaking around $10$ km s$^{-1}$, similar to our results. Very recently, \citet{2025ApJ...988..183P}, report a $23$\% runaway fraction above $25$ km s$^{-1}$, out of 104 stars from the original bow shock catalogs used in our work. The differences in the exact values are unsurprising given the difference in correction for Galactic rotation: the different approaches return different peculiar angular motions per star (e.g., Figure \ref{fig:method}), but also lead to different final samples: for example, we remove $20$ targets where too few neighbors where available with sufficient \textit{Gaia} data quality. 

As noted already by \citet{2022AJ....164...86K}, our runaway fractions are consistent with measurements made for stars of similar type without bow shocks \citep[see also, e.g.,][]{2023A&A...679A.109C,2023A&A...670A.108S, 2025A&A...694A.250C}. From searches of runaway systems, it is evident that the opposite is true as well: the majority of runaway stars does not drive a detected bow shock \citep{peri2012,peri2015}. As also argued by \citet{2022AJ....164...86K} and \citet{2025A&A...698A..64R}, based on studying the kinematics and the environment of individual bow-shock driving stars in detail, this shows how the ISM properties are the key driver of bow shock formation and detection. This result mimics what is known for, for instance, X-ray binary studies, where a very low percentage of systems shows feedback nebulae from accretion-driven feedback \citep[see, e.g., recent discussions in][]{2025A&A...696A.223A,2025A&A...696A.222M}. 

\subsection{Galactic rotation corrections}

One specific point to focus on, is the effect of systematic biases in $\mu_{l*}$ as a function of $l$. If residuals remain in the corrected $\mu_{l*, \rm corr}$, these likely average to zero across the full range of $l$. However, as massive stars are concentrated within the Solar circle, our systems are clustered in the first and fourth Galactic quadrants, where $\mu_{l*}$ due to Galactic rotation is predominantly negative (compared of predominantly positive in the second and third quadrants). We find that the residuals after Galactic rotation in the first and fourth quadrant leave a mean negative $\mu_{l*, \rm corr}$, while $\mu_{b, \rm corr}$ averages out to zero. When converting to the direction of peculiar motion in the plane of the sky -- which can be compared to observed bow shock position angles -- this bias results in non-zero mean motion in \textit{both} right ascension and declination. Any calculations based on these biased measurements will (i) lead to overestimated velocities and (ii) incorrect position angles. Performing our full analysis with case A of our Galactic rotation corrections, for instance (see Section \ref{sec:astrometry}), implies ISM velocities $\sim 10$ times larger than we find in our actual analysis: when peculiar velocities are larger, the ISM needs to move faster to mis-align a bow shock. 

There are different potential approaches to correct for this bias. For instance, one could find a combination of the Oort constants $A$ and $B$ that returns the same combination $A-B$, but minimizes the average $\mu_{l*,\rm corr}$ and $\mu_{b,\rm corr}$. The same approach could be taken with $(U,V,W)_\odot$. However, firstly, it becomes an unconstrained problem, with two minimized values but five parameters. Secondly, while this approach optimizes the average corrected angular motions across the entire sample, there is no reason to assume that this minimization corresponds to a rotation curve that accurately describes the local motion for each individual object. Possibly, the average peculiar velocities across the sample will be of the correct order of magnitude, but the peculiar motion directions and therefore misalignment angles will be unreliable. Furthermore, the systematic uncertainty with this approach is difficult to assess, and would dominate greatly over the statistical \textit{Gaia} uncertainties on parallax and proper motion. 

\citet{2022AJ....164...86K} instead take an approach that is similar, but optimizes a different metric. Specifically, their analysis finds $(U,V,W)_\odot$ that ensures the stellar kinematics are uncorrelated with Galactic longitude. This analysis approach leads to a high probability that the bow shock orientations and stellar peculiar motion directions are drawn from the same distribution. As we focus on modeling the distribution of $\beta$, we cannot use the same approach in our Galactic rotation correction. 

Given these challenges with using Galactic rotation models \citep[e.g.,][]{1987pgim.book.....S,2007A&A...467L..23C}, we instead opt to use the close neighbors of each star. This method is significantly more expensive in terms of computation and catalog fetching. It furthermore leads to significantly larger uncertainties in the corrected stellar motion. However, these are not dominated by -- potentially difficult to quantify -- systematic uncertainties in the rotation curve. Asymmetric drift, e.g., the effect that the motions of stars in the Milky Way depend on their age \citep[e.g.,][]{2008gady.book.....B}, may have an effect on our measurements if the majority of good quality neighbors has a different age. However, we measure sample-average $\mu_{l*,\rm corr}$ and $\mu_{b,\rm corr}$ values of zero within uncertainties. Therefore, we believe that this effect at most acts to increase the uncertainty of our peculiar angular motions, leading to larger errors on $v_{\rm star}$, $\beta$ and more conservative estimates on ISM motion. 

Finally, it is noteworthy that the bias in Galactic rotation correction is visible in Figure \ref{fig:method} because we discuss a sufficiently large sample. In analyses of individual objects, this effect may not have been noticed. For instance, \citet{2024MNRAS.532.2920V} investigate the bow shock of LS 2355\footnote{Discovered by \citet{fermipaper2018} and therefore not part of our sample.} and find a low peculiar velocity of only $7.0\pm2.5$ km s$^{-1}$. The system is moving towards (and partially embedded in) an HII region, implying that the peculiar movement may be accurate and that movement of the HII region increases the relative velocity difference. Alternatively, the magnitude of the peculiar velocity may be underestimated. Whichever of these is the case, we encourage caution when studying individual bow shocks where the overall statistics of a sample is not available. 
 
\begin{figure}
\includegraphics[width=\columnwidth]{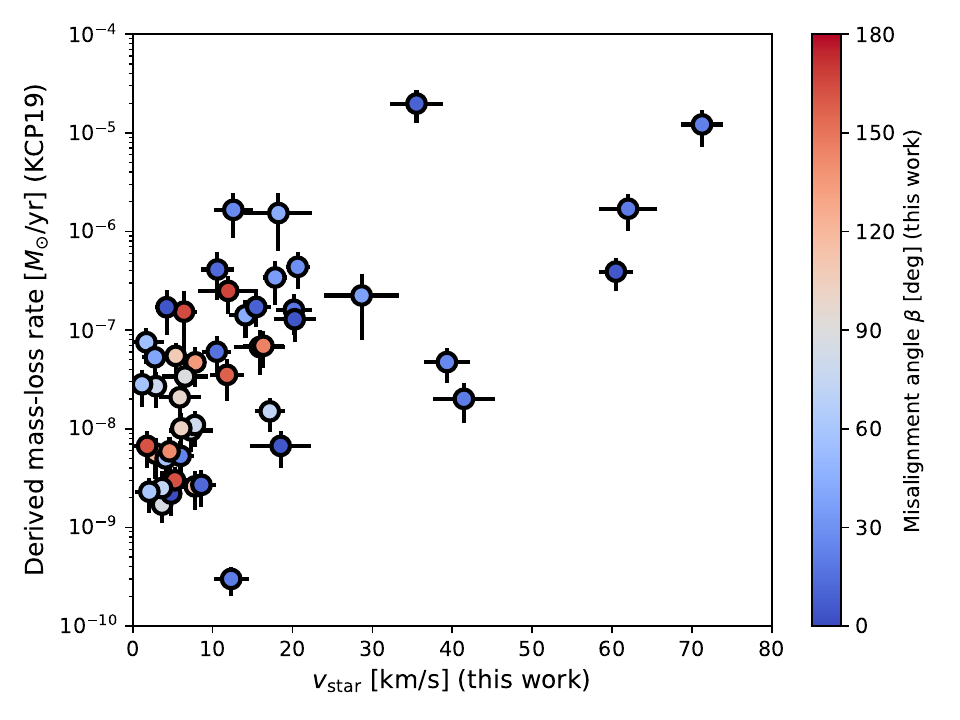}
 \caption{The stellar wind mass-loss rates derived by \citet{2019AJ....158...73K} versus the stellar peculiar velocity (this work) color coded by $\beta$ (this work). A broad correlation between mass loss rate and velocity can be seen, which we deem to be consistent with underestimates of the derived mass-loss rates at low velocities. The largest mis-alignment angles are also seen for the lowest derived mass-loss rates.}
\label{fig:mdots}
\end{figure}

\subsection{Implications for ISM motion, massive star winds, and bow shock physics}

In our analysis, we find that bulk ISM motions of the order $10-15$ km s$^{-1}$ can explain the distribution of bow shock misalignment angles for our entire sample. What could explain this magnitude of bulk motion? One possibility is noted by \citet{2025A&A...698A..64R}, discussing deviations from circular motion of the ISM. For instance, spiral arm perturbations may lead to peculiar gas motion in radial (towards the Galactic center) and azimuthal (against Galactic rotation) direction \citep{1969ApJ...158..123R,2000RvMA...13...97E,2018MNRAS.474.2028R}. The typical magnitude of these motions are $\sim 10$ km s$^{-1}$ \citep{2012ApJ...754...62A,2015A&A...579A..91W}, but may be sometimes be significantly higher \citep{1993A&A...275...67B}. Measurements of the kinematics of atomic hydrogen within the Solar circle also reveal similar motions: \citet{2017A&A...607A.106M}, for instance, report velocity dispersion compared to circular motion of $8.9\pm1.1$ km s$^{-1}$. We deem it likely that the orientations of massive star bow shocks probe the same motion of the ISM, unsurprisingly finding approximately the same required ISM bulk velocity. 

It is important to stress that bow shocks are sensitive to the bulk motion of the ISM on scales of parsecs (e.g., arc minutes). In comparison, mapping studies of the entire Milky Way inevitably tend to have larger resolution. The consistency between our velocity magnitude and larger scale measurements, implies that we find no evidence for significantly different ISM kinematics on smaller scales. This holds despite the difference between the type of bulk ISM flow models: we use a non-zero mean velocity offset from the LSR, instead of a velocity dispersion centered around the LSR motion. We postulate that this difference is the result of our sample selection: being a bow shock sample, it is more sensitive to ISM motions above $10$ km s$^{-1}$, where even stationary massive stars experience a supersonic velocity contrast with the ISM. Instead, in regions where the ISM motion deviates less from the LSR, only \textit{classic} bow shocks are expected. As a result, our sample probes the higher velocity part of the Galaxy-wide ISM velocity dispersion. 

When we focus specifically on systems facing HII regions, and assume the ISM motion is dominated by outflows from these regions, we find bulk velocities of at least $25$ km s$^{-1}$. Such larger velocities are not surprising in the context of flows from HII regions, in particular in the light of the `champagne' model \citep{1979A&A....71...59T, 1979ApJ...233...85B}. Velocities of the order seen in our analysis are consistent with the expectation for these outflows \citep[e.g.,][]{2002ApJ...580..969S} and match observed flows velocities of individual HII regions \citep[e.g.,][]{2014A&A...563A..39I}. More generally, expanding HII regions may be the origin of these specific ISM motions \citep{1990ApJ...349..126F, 1996ApJ...469..171G, 2002ApJ...580..980K}; however, the expansion velocities of HII regions are observed at slightly lower values \citep[e.g.,][]{2025ApJ...990...30F}. Not all massive stars close to HII regions may experience this enhanced ISM motion, due to asymmetry of outflows or the evolutionary state of the HII region. However, we postulate that we observe this effect here because the bow-shock selected sample is biased towards the massive stars where this effect occurs: without the HII region's outflow (or ISM motion more generally), the majority of these stars ($\sim 60\%$) is subsonic and would not form a bow shock. 

For the ISM motions calculated in our work, we also investigate the effect on the stand-off distance via the parameter $f_{\rm corr}$. Overall, we find that $f_{\rm corr}$ peaks at a value of $1$, as can be seen in the blue and red curves in the right-hand panel of Figure \ref{fig:results_per_env}. However, the distribution of $f_{\rm corr}$ is evidently very broad, spanning more than four orders of magnitude for the isolated and non-isolated case. The origin of this broad distribution is the grouping of different peculiar velocities: the analysis per velocity quintile (Figure \ref{fig:results_per_v}) shows how much the $f_{\rm corr}$ distribution depends on the typical peculiar stellar velocities in the sample. In particular for slow systems with sub-sonic velocities, the wind parameters can therefore easily be underestimated by a factor few to more than an order of magnitude. As we perform a statistical analysis, we cannot provide correction factors for individual objects. However, the mode and width of the $f_{\rm corr}$ distributions (the latter even at supersonic velocities) show how ISM motion is a dominant source of uncertainty when converting bow shock properties to stellar wind properties. 

To further highlight the latter finding, we plot our stellar peculiar velocities in Figure \ref{fig:mdots}. We compare these measurements with the mass-loss rates derived from the stand-off distances of these bow shocks by \citet{2019AJ....158...73K}. A broad correlation between the two quantities is observed. Some of this correlation could be explained by a dependence of both mass-loss rate and runaway fraction on stellar type. Otherwise, a role may be played by detectability: for large peculiar velocities, low mass-loss rates would lead to very small stand-off distances and small bow shocks (in angular terms, and in the absence of ISM motion), potentially making them harder to detect. However, while the peculiar velocities in this diagram span barely more than an order of magnitude, the mass-loss rates differ by more than four orders of magnitude. Therefore, we postulate that this broad correlation could for a significant part be caused by the effect of ISM motion: at low $v_{\rm star}$, $f_{\rm corr}$ is larger then at higher stellar velocities, leading to systematic underestimate of the inferred $\dot{M}_{\rm wind}$. The effect of ISM motion should therefore be taken into account explicitly when applying the relatively under-explored bow shock method of measuring stellar wind mass loss rates \citep{vink2022,2025AJ....169..337W}. 

That statement raises the question how our results can be applied in future studies or can be expanded upon. Our analysis shows that, if the stellar wind parameters and ISM density are known independently, the motion of the ISM can be determined uniquely from the observed bow shock orientation and stellar peculiar velocity. Similarly, if the orientation of the bulk ISM motion is known, one can in principle determine its velocity and thereby $f_{\rm corr}$. Such a measurement would then improve inferences of the stellar wind parameters from the stand-off distance and ISM density. An important caveat, however, is that the misalignment angle show asymptotic behavior $\beta \rightarrow \alpha$ for large $v_{\rm ISM} / v_{\rm star}$. Therefore, for bow shocks orientated towards HII regions, where $\alpha \approx \beta$, only a lower limit on the ISM velocity can be determined. This leads to a lower limit on $f_{\rm corr}$, which still improves upon stellar wind inferences that ignore ISM motion. A better case would be a misaligned bow shock close to an HII region, but not directly pointed towards it; however, the more the bow shock points towards the HII region, the stronger the argument becomes that the ISM motion is dominated by the HII region's outflow. 

Another avenue of future research is to revisit the effect of 3D motion on this problem. We treat the current problem in 2D due to the availability of reliable measurements of motions in the plane of the sky, while radial velocities are not known for all systems. As mentioned earlier, the considered bow shocks are all likely seen at relatively high inclination (e.g., indicating motion preferentially in the plane of the sky), as otherwise their projected shapes deviate significantly from the arcuate shapes used to select them \citep{2016MNRAS.456..136A,2018MNRAS.477.2431T}. In the presence of ISM motion, this same argument holds, however, now considering the direction of the effective (net) velocity difference between star and ISM in 3D. Ignoring this effect may, however, lead to underestimates of the total magnitude of ISM velocity. New numerical simulations of bow shock morphology and detectability \citep[building on, e.g.,][]{2016MNRAS.456..136A} for various observing angles and 3D angles between ISM and stellar motion would help to quantify this effect. 

\section{Acknowledgments}
The authors are grateful to the referee for their rapid and constructive review of this work. The authors thank Mitchel Stoop for useful discussions on the \textit{Gaia} analysis. This work would not have been possible without the significant efforts of previous authors to make the data underlying their papers easily and publicly available through Vizier. JvdE was supported by funding from the European Union's Horizon Europe research and innovation programme under the Marie Skłodowska-Curie grant agreement No 101148693 (MeerSHOCKS). This research has made use of the VizieR catalogue access tool, CDS, Strasbourg, France. This research has made use of the Astrophysics Data System, funded by NASA under Cooperative Agreement 80NSSC21M0056. This research has made use of the SIMBAD database, operated at CDS, Strasbourg, France. This work made use of Astropy: \footnote{https://www.astropy.org} a community-developed core Python package and an ecosystem of tools and resources for astronomy \citep{astropy:2013, astropy:2018, astropy:2022}. This work has made use of data from the European Space Agency (ESA) mission {\it Gaia} (\url{https://www.cosmos.esa.int/gaia}), processed by the {\it Gaia} Data Processing and Analysis Consortium (DPAC, \url{https://www.cosmos.esa.int/web/gaia/dpac/consortium}). Funding for the DPAC has been provided by national institutions, in particular the institutions participating in the {\it Gaia} Multilateral Agreement.

\section*{Data Availability}
A \textsc{GitHub} reproduction repository will be made public upon formal acceptance of this paper at \url{https://github.com/jvandeneijnden/MisalignedBowshocks}, containing all files to repeat the data selection and analysis presented in this paper. A stable release of this reproduction package will also be released via \textsc{zenodo} via the following DOI: \url{doi.org/10.5281/zenodo.21903643}.

\bibliographystyle{mnras}
\bibliography{main.bib}


\bsp	
\label{lastpage}
\end{document}